%% file: SR_collection.tex
\documentclass[sigconf]{acmart}
\usepackage{color} 
\usepackage{xspace} 
\usepackage{booktabs}
\usepackage{enumitem}

\AtBeginDocument{%
  \providecommand\BibTeX{{%
    \normalfont B\kern-0.5em{\scshape i\kern-0.25em b}\kern-0.8em\TeX}}}

\newcommand*\flabel[1]{\colorbox{black}{\textcolor{white}{#1}}}
\copyrightyear{2022}
\acmYear{2022}
\setcopyright{acmcopyright}
\acmConference[SIGIR '22]{Proceedings of the 45th International ACM SIGIR Conference on Research and Development in Information Retrieval}{July 11--15, 2022}{Madrid, Spain.}
\acmBooktitle{Proceedings of the 45th International ACM SIGIR Conference on Research and Development in Information Retrieval (SIGIR '22), July 11--15, 2022, Madrid, Spain}
\acmPrice{15.00}
\acmISBN{978-1-4503-8732-3/22/07}
\acmDOI{10.1145/3477495.3531748}

\usepackage{caption}
\usepackage{subcaption}
\usepackage{multirow}
\begin{document}
\fancyhead{}

\title{From Little Things Big Things Grow: A Collection with Seed Studies for Medical Systematic Review Literature Search}

%
%
%
%
%

\author{Shuai Wang}

\affiliation{%
	\institution{The University of Queensland}
	\city{St Lucia}
	\country{Australia}
}
\email{shuai.wang2@uq.edu.au}

\author{Harrisen Scells}
\affiliation{%
	\institution{The University of Queensland}
	\city{St Lucia}
	\country{Australia}
}
\email{h.scells@uq.net.au}

\author{Justin Clark}
\affiliation{%
	\institution{Bond Institute for Evidence-Based Healthcare}
	\city{Gold Coast}
	\country{Australia}
}
\email{jclark@bond.edu.au}

\author{Bevan Koopman}
\affiliation{%
	\institution{CSIRO}
	\city{Herston}
	\country{Australia}
}
\email{bevan.koopman@csiro.au}

\author{Guido Zuccon}
\affiliation{%
	\institution{The University of Queensland}
	\city{St Lucia}
	\country{Australia}
}
\email{g.zuccon@uq.edu.au}

\vspace{5pt}
\begin{abstract}
Medical systematic review query formulation is a highly complex task done by trained information specialists. Complexity comes from the reliance on lengthy Boolean queries, which express a detailed research question. To aid query formulation, information specialists use a set of exemplar documents, called `seed studies', prior to query formulation. Seed studies help verify the effectiveness of a query prior to the full assessment of retrieved studies. Beyond this use of seeds, specific IR methods can exploit seed studies for guiding both automatic query formulation and new retrieval models. One major limitation of work to date is that these methods exploit `pseudo seed studies' through retrospective use of included studies (i.e., relevance assessments). However, we show pseudo seed studies are not representative of real seed studies used by information specialists. Hence, we provide a test collection with real world seed studies used to assist with the formulation of queries. To support our collection, we provide an analysis, previously not possible, on how seed studies impact retrieval and perform several experiments using seed-study based methods to compare the effectiveness of using seed studies versus pseudo seed studies. We make our test collection and the results of all of our experiments and analysis available at \url{http://github.com/ielab/sysrev-seed-collection}.
\end{abstract}


\maketitle

\input{introduction}

\input{relatedwork}

\input{testcollection}

\input{Query-Formulation}

\input{Seed-driven-Document-Ranking}

\input{snowballing-baselines}

\input{conclusion}

\section{Acknowledgment}

Shuai Wang is supported by a UQ Earmarked PhD Scholarship and this research is funded by the Australian Research Council Discovery Projects programme ARC DP DP210104043.


\graphicspath{ {./graph/} }
\clearpage
\bibliographystyle{ACM-Reference-Format}
\bibliography{bibliography}

\appendix

\end{document}

%% file: introduction.tex
\section{Introduction}

Systematic reviews are comprehensive reviews about a particular body of literature. Systematic reviews are used extensively in medicine for various reasons: by clinicians to make health decisions and by governmental and institutional policy and practice decisions about health topics. To ensure comprehensiveness, creators of systematic reviews employ a standardised approach to searching for literature in the form of Boolean queries. However, creating Boolean queries is often so complex that the authors of systematic reviews often engage with information specialists: highly specialised librarians skilled in searching for medical literature for systematic reviews.
Boolean query could be considered one of the most important aspects of a systematic review: it not only controls how many potentially relevant studies are retrieved but, perhaps more importantly, the total number of studies to screen. In other words, because systematic reviews aim to be comprehensive, the researchers will screen (i.e., assess for relevance to be included in the final systematic review) every study retrieved by the Boolean query. To ensure that no relevant studies are left out in the screening phase and to account for assessor biases, it is common for at least two assessors to screen studies. Naturally, since there may be thousands, if not tens of thousands of studies to screen, and since the screening process is duplicated by independent assessors, the Boolean query must be as effective as possible.

One technique that information specialists may use to assist in creating effective queries is to use `seed studies': exemplar documents known to the information specialists a priori and often provided by the researchers of the systematic review. Information specialists use seed studies in various ways: identifying potential terms or phrases to use in the Boolean query; validating their search by analysing the precision and recall of their search against seed studies. Although, as we will demonstrate in the following sections, seed studies may not be a good indicator for query effectiveness.

The techniques that information specialists use to design and validate their search using seed studies have not gone unnoticed by the Information Retrieval community. There have been several studies that have exploited seed studies for automatic query formulation~\cite{scells2021comparison} and screening prioritisation (i.e., ranking the set of retrieved studies)~\cite{lee2018seed, wang2022seed}. However, one major limitation of these previous studies is that the collections that they used \textit{did not contain any seed studies}. 
Instead, a small portion of the final relevant studies were held out as a set of `pseudo seed studies'. 
Given that these works use studies that are known to be relevant and included in a systematic review, we believe that the effectiveness of these methods is overestimated. 
However, methods that use pseudo seed studies are relying of documents that are known to be relevant and included in a systematic review; we posit and show that the effectiveness is thus overestimated. 
To address this issue, this resource paper presents a new Information Retrieval test collection for systematic review literature search; our novel contribution is that our collection includes real seed studies for each systematic review topic.

Our test collection enables the development of new Information Retrieval methods for systematic review literature search that depend on or utilise seed studies. To this end, we also investigate the impact of seed studies versus pseudo seed studies by reproducing two prominent methods from the literature (automatic query formulation and seed-driven document ranking). In addition, we demonstrate a new technique that can use seed studies which is often actually utilised when constructing a systematic review but overlooked in Information Retrieval methods: snowballing (otherwise known as citation chasing). This method is discussed in more detail in the sections that follow. 
%
%
We make our test collection and the results of all of our experiments and analysis available at \url{github.com/ielab/sysrev-seed-collection}.

\begin{figure*}
	\vspace{-8pt}
	\includegraphics[trim=0 142px 270px 0,clip,width=\textwidth]{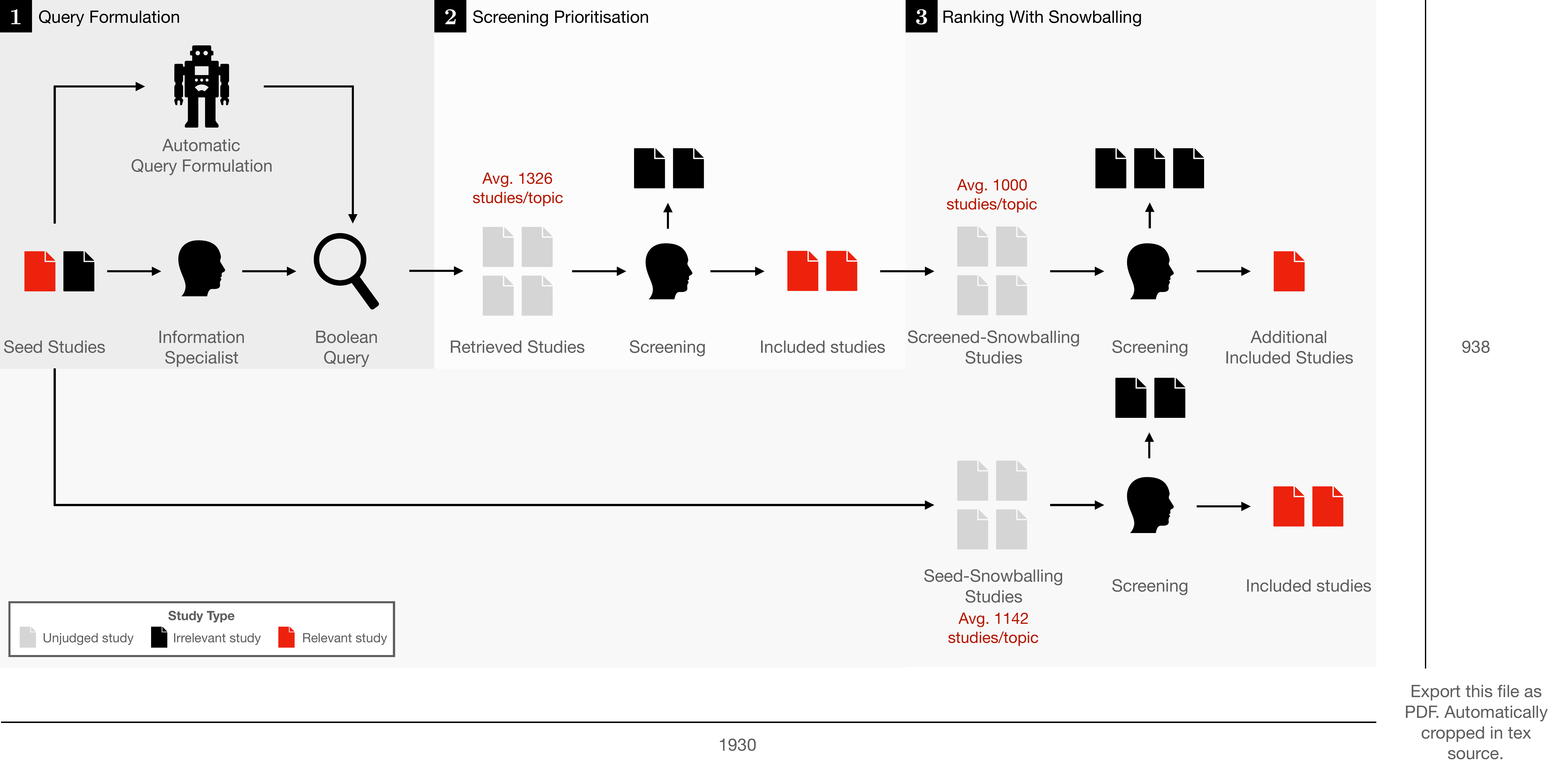}
	\caption{High-level overview of the systematic review creation processes that are relevant to our collection, and several use cases of our collection for automating these processes. Also shown are the three use-cases we will demonstrate for our collection: \flabel{1} query formulation, \flabel{2} screening prioritisation, and \flabel{3} ranking with snowballing.}
	\vspace{-12pt}
	\label{fig:process}
\end{figure*}

%% file: relatedwork.tex
\vspace{-16pt}
\section{Background \& Related Work}
\label{sec:related-work}

\subsection{Systematic Review Creation}

Creating a systematic review is a highly time-consuming, costly and complex task, which involves numerous stages~\cite{Bullers:2018vc}. One of the initial stages in the process is searching and screening medical literature. This stage contributes the most to the costs of a systematic review~\cite{McGowan:2005up}, and it is at this stage that our collection is intended to be used. Figure \ref{fig:process} provides a high-level overview of the early stages of the systematic review process, and the possible uses cases or tasks that our collection enables. The figure shows that information specialists first use seed studies to create a Boolean query. The most common use for seed studies is identifying and extracting terms that are topically related to the systematic review. The Boolean query is then executed to retrieve a set of studies for manually screening by the researchers of the systematic review. In reality, there are two phases of screening: one at the abstract level, which we have described here; and one at the full text level, where the full-text of the abstract level studies are assessed for inclusion in the systematic review. In our collection, the data that we were provided did not contain the abstract level relevance information of studies, nor the studies that were marked as not-relevant at an abstract level and full text level. We will discuss the implications of this in more detail in the sections that follow. 
The stages that follow in the systematic review creation process (e.g., data extraction or study synthesis) are not relevant to this collection, but are impacted by the methods that are created using our methods. For example, one of the use-cases of our collection, screening prioritisation (i.e., ranking studies), can be used to begin the following stages earlier

\vspace{-8pt}
\subsection{Seed Studies}
Seed studies are used extensively by information specialists for query formulation and validation in systematic review literature search~\cite{suhail2013methods,hausner2012routine}. The use of seed studies is somewhat related to various explicit relevance feedback mechanisms, e.g. Rocchio~\cite{rocchio1965information}. In the Information Retrieval domain, explicit relevance feedback has been used in applications such as query expansion~\cite{lavrenko2017relevance,kaptein2008impact,colace2011improving}, active learning~\cite{pereira2020iterative}, and user modelling~\cite{jayarathna2015unified}.

Active learning using explicit relevance feedback has been extensively studied in the area of technology assisted reviews~\cite{cormack2016scalability,zhang2018effective,cormack2014evaluation,cormack2015autonomy,lagopoulos2018learning,lee2018seed,zou2018technology,abualsaud2018system}: a catch-all phrase for high-recall tasks such as systematic review literature search, legal search, patent search, among others.
The difference between explicit relevance feedback and seed studies is subtle but important. Seed studies cannot be considered relevant because they (1) have not necessarily been assessed as such when provided to an information specialist; and (2) may not be studies suitable for a systematic review (i.e., systematic review typically synthesise the results of several \textit{randomised controlled trials}).\footnote{Randomised controlled trials with many patients are considered a higher form of evidence compared to studies like case reports which may pertain to a single patient.} Therefore, although seed studies may be used in the place of explicit relevance feedback documents for certain applications such as query expansion, they should not be considered of equal weight in terms of relevance to a systematic review. 

\vspace{-8pt}
\subsection{IR Systematic Review Collections}
\label{sec:related.ir}
There are several test collections for research in systematic review literature search. These include i) the CLEF Technology Assisted Review collections~\cite{kanoulas2017clef,Kanoulas2018CLEF2T,kanoulas2019clef}; ii) a similar test collection from~\citet{scells2017test}; iii) a test collection about systematic review updates~\cite{alharbi2019dataset}; and iv) collections for tasks such as data extraction and summarisation~\cite{norman2018data}. However, as we have raised earlier, none of these collections contain seed studies. This means that the true effectiveness is unknown of any existing methods that claim to use seed studies. Our collection enables the proper evaluation of existing and future methods that use seed studies.

\vspace{-8pt}
\subsection{Snowballing}
In order to maximise comprehensiveness, one additional step in the systematic review creation process that is currently unexplored by current Information Retrieval methods is that of snowballing or citation chasing~\cite{snowballing-definition-choong}. Snowballing identifies new studies to screen from the references or citations of other relevant studies. Snowballing is often performed `backwards' and `forwards' over the references to identify all studies that cite a given study and all studies cited by a given study. 
Although commonly performed in practise, there are very few studies in the literature that explore the topic of snowballing~\cite{animesh2018neuralparscit}. Nevertheless, snowballing has been shown to contribute 51\% of the included studies for a systematic review~\cite{greenhalgh2005effectiveness}.
The reason we investigate snowballing is to provide a more complete set of relevance assessments. Our preliminary experiments showed that not all included studies were retrieved by the Boolean query. We obtained total recall for many of our topics only when snowballing. Therefore, we perform snowballing to collect a more realistic set of studies that would be screened by researchers of a systematic review.

%% file: testcollection.tex
\section{Collection Details}
\label{sec:collection}

\begin{table}[t]
	\centering
	\begin{tabular}{p{70pt}p{154pt}}
		\toprule
		\bf Name & \bf Description\\\midrule
		ID&Unique ID of a topic.\\
		Link to Review&Link (URL) to the published review.\\
		Title&Title of the systematic review.\\
		Description & Topical summary of the search. \\
		Date restrictions&Date restriction used to retrieve studies. \\
		PubMed query&PubMed query used to retrieve studies.\\
		Seed studies&PMIDs of seed studies.\\
		Included studies&PMIDs of included studies.\\
		Retrieved studies&PMIDs of studies retrieved by the query.\\
		Snowballed studies&PMIDs of snowballed studies.\\
		\bottomrule
	\end{tabular}
	\caption{Attributes of each topic in our collection. PMID refers to `PubMed identifier', and is used to uniquely refer to a study or document in the PubMed database.}
	\label{table:column}
	\vspace{-26pt}
\end{table}


\subsection{Topic Attributes}
The basis of our collection comes from our co-author at the Bond Institute for Evidence-Based Healthcare, who is a senior information specialist. Each topic is a systematic reviews created using Boolean queries developed by our co-author over the past five years. We were provided unstructured data that we organised into 40 topics. 

Each topic in our collection has several attributes, as shown in Table~\ref{table:column}.
Firstly, we assign a unique ID to each systematic review in our collection. As our collection contains completed and published systematic reviews, we also include the title and the URL of the published review. We also include a short description of the search provided by our information specialist co-author.

Apart from these metadata attributes, each topic also contains information about the query. Specifically, the PubMed Boolean query, the date restrictions applied to the search (i.e., used to restrict which studies are retrieved based on the publication date), and the seed studies used to assist in the development of the query. Note that it is common to use multiple search queries across several medical literature databases to maximise recall. However, most of these databases require a monetary subscription to access. Therefore, we choose to only include the PubMed query in our collection. This is a limitation of all collections listed in Section~\ref{sec:related.ir}, including ours. 

The next set of attributes corresponds to the relevance assessments. The included studies represent studies that are relevant at the full-text level. This means that these studies were assessed as sufficient to be included in the final review. As such, it should be noted that evaluation of an Information Retrieval system using our collection is more `strict' (i.e., it is naturally more challenging to identify those studies that will be included in the final systematic review than those that are potentially relevant at an abstract level). Additionally, we provide the set of retrieved studies that were retrieved by the Boolean query.

The final attribute of each topic corresponds to snowballed studies. We include two sets of snowballed studies for each topic: one corresponding to snowballing the seed studies and another corresponding to snowballing the retrieved included studies. The second set of snowballed studies corresponds to simulating the process researchers of a systematic would do in order to identify additional relevant studies after assessing the set of retrieved studies. Note that this is an overlooked process in Information Retrieval methods in this space. We take this opportunity to do an initial investigation of the impact snowballing has on retrieval effectiveness and integrate the comparison of snowballing seed studies into the investigation.

\begin{figure*}[t!]
	\centering
	\begin{subfigure}[h]{0.662\textwidth}
		\includegraphics[width=\textwidth]{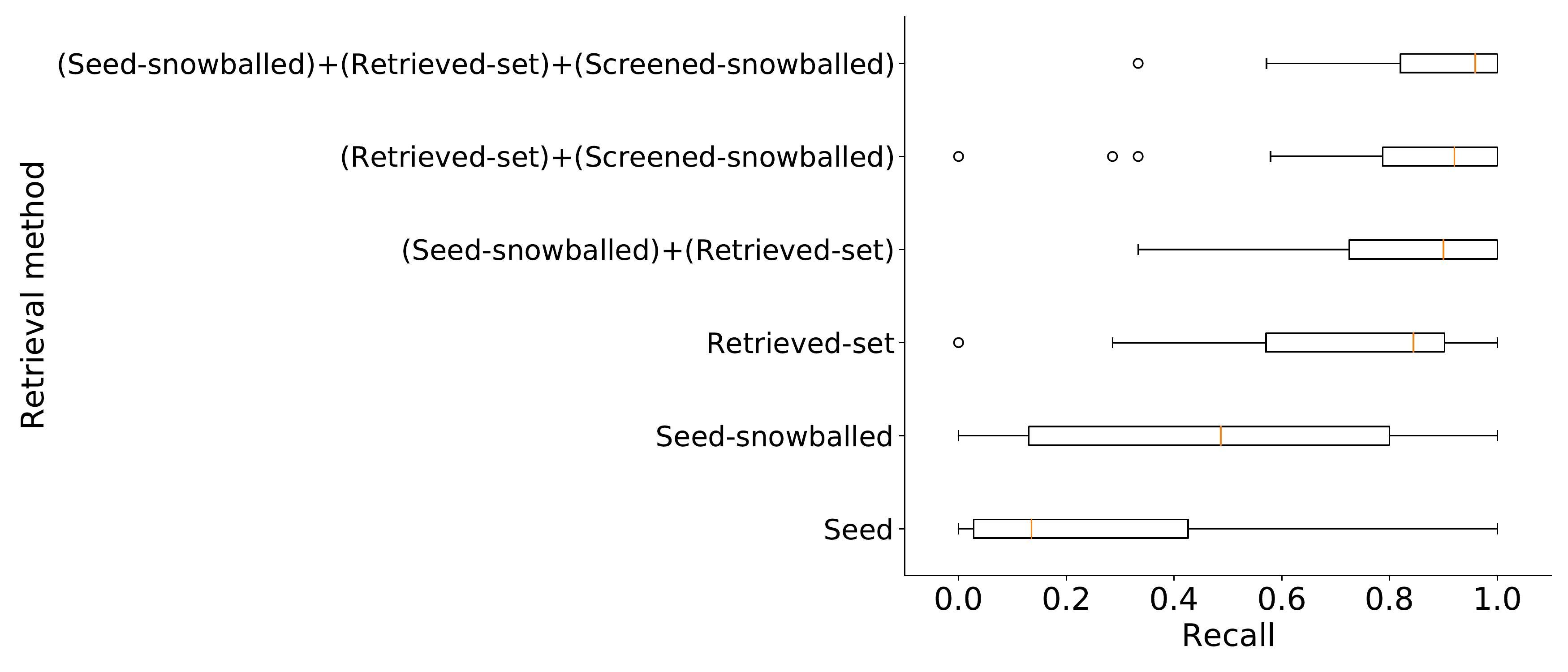}%
		\label{fig:a}%
	\end{subfigure}
	\hfill
	\begin{subfigure}[h]{0.332\textwidth}
		\includegraphics[width=\textwidth]{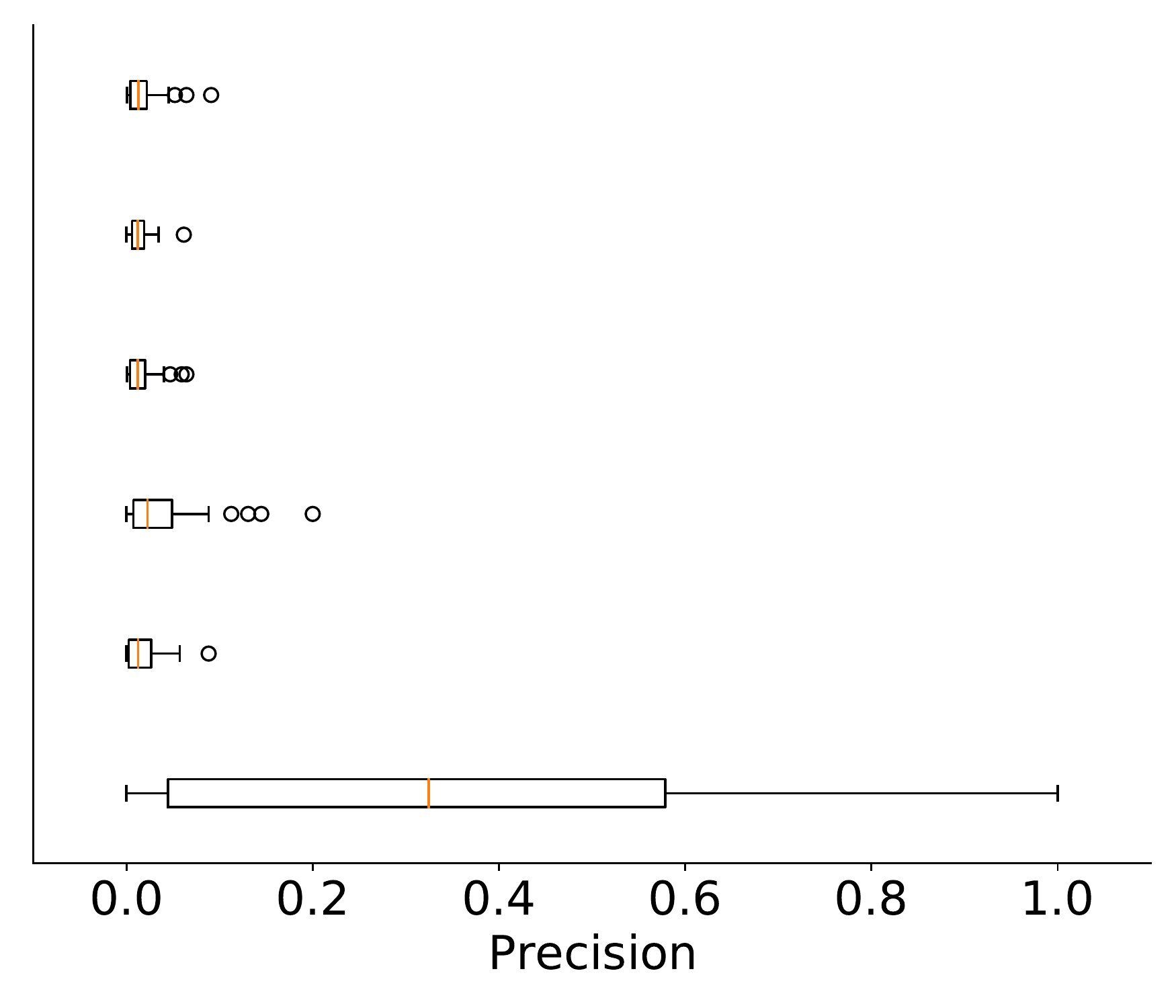}%
		\label{fig:b}%
	\end{subfigure}
	\caption{Recall and precision box plots of included studies for different retrieval methods. The methods listed include seed studies (seed), the Boolean query results (retrieved set), the two snowballing sets (seed-snowballing, i.e., snowballing applied to seed studies; and screened-snowballing, i.e., snowballing applied to retrieved included studies). These sets of studies are also combined in numerous ways, as indicated by `+'.}
	\label{figure:precision_recall_boxplot}%
	\vspace{-12pt}
\end{figure*}
\begin{figure}[t!]
	\includegraphics[width=\linewidth]{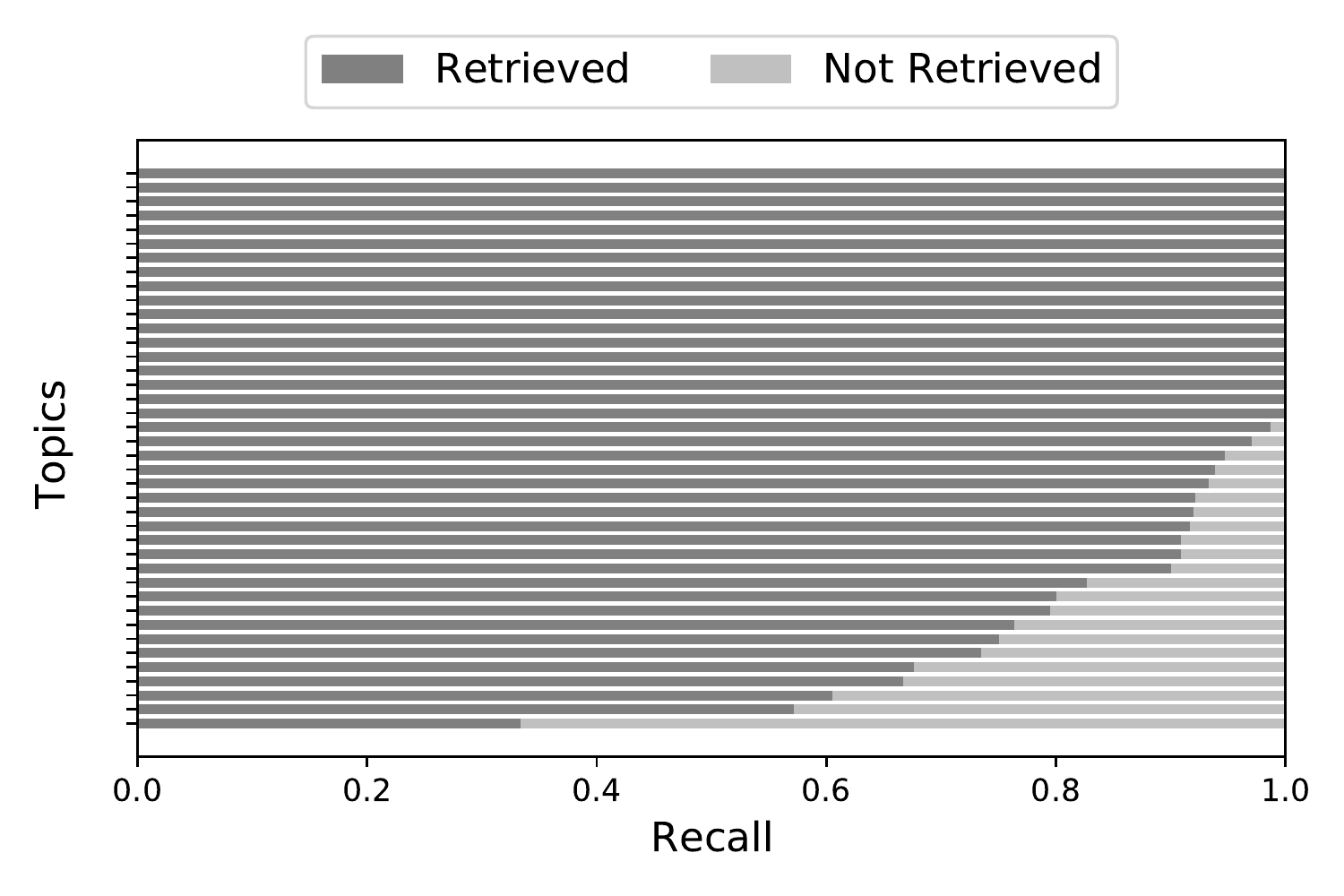}
	\caption{Recall distribution of all of the topics given the combined set of studies that includes retrieved studies, the seed-snowballed set, and the screened-snowballed set.}
	\label{fig:bar_chart}
	\vspace{-12pt}
\end{figure}

\subsection{Data Processing}

Several attributes of topics required data processing beyond extraction. Specifically, the included studies, the retrieved studies, and the snowballed studies required additional processing to generate.

\subsubsection{Included studies}
The raw data that we were provided did not include any relevance assessment information. As such, we were required to create the relevance assessments for each topic. To do this we manually extracted the list of included studies that were used in the analysis of the systematic review. Note that this was not as simple as scraping the references of a systematic review, but involved manually matching the citations used in the analysis to references of published studies. Sometimes an included study is not available in PubMed. Here, we do not include these studies in our relevance assessments as we assume it is not retrievable.

\subsubsection{Retrieved studies}
The raw data that we were provided also did not include the studies that were retrieved by the Boolean query. As such, we reproduced the search for all topics to obtain a set of retrieved studies. We automated this step by using the Entrez API~\cite{maglott2005entrez}. The date restrictions were applied to each Boolean query, therefore others should be capable of reproducing the set of retrieved documents for each topic if necessary.

\subsubsection{Snowballed studies}
Rather than manually snowballing studies, we use two tools: Citationchaser~\cite{haddaway_n_r_2021_4543513} and SpiderCite~\cite{2-week-sys-review}. We first use CitationChaser to format a list of studies into the specific format required by SpiderCite.\footnote{The format is RIS; a standardised tag format that enables citation programs to exchange data, and the only format SpiderCite supports as input.} Using SpiderCite, we obtain the list of cited and citing studies given the input set. Each study in SpiderCite has a DOI that we use to retrieve the study from PubMed. 

%

\subsection{Collection Statistics \& Analysis}


\subsubsection{Seed and included studies analysis}
On average, 15 seed studies are used for query construction, with a median number of 12. The average number of included studies per systematic review is 29, with a median number of 11.5.
When comparing the overlap of seed studies and included studies, we found that, on average, only 36.3\% of seed studies are contained in the included studies, and make up 26.1\% of all included studies. This finding suggests that even though seed studies are helpful for Boolean query construction, most are disregarded after the query construction phase. This demonstrates that treating included studies as pseudo seed studies as in prior work~\cite{lee2018seed, scells2021comparison} may lead to inaccurate results. 

%

\subsubsection{Searching analysis}
\label{searching analysis}
When using the Boolean queries and date restrictions to retrieve studies, the mean number of retrieved documents per query is approximately 1,326, with median 709. Across several topics, we also found that not all included studies could be found in the retrieved studies. 
We found that only 75.5\% of all included studies across all topics can be found inside the retrieved studies. There are two possible reasons why some included studies can not be found in candidate documents. The first is that some included studies may have been identified through snowballing. The second is that some studies may have been identified by searching in a database other than PubMed (although such studies may still exist in PubMed).

However, topic \#18 does not contain any included studies in the retrieved studies. This particular systematic review only has a single included study. For topic \#18, the reason that only one study is included is because many studies screened as relevant at the abstract level encountered a high risk of bias in the full-text.

\subsubsection{Snowballing analysis}
Our collection includes two snowballed sets of studies for each topic. The first set corresponds to snowballed seed studies (\textbf{seed-snowballing}) and the other corresponds to simulated snowballing of the included studies in the retrieved set (\textbf{screened-snowballing}).

For the seed-snowballing set, we find that 35 topics retrieved at least one included study. The topics that did not retrieve any included studies using this snowballing technique were 46, 52, 53, 66, and 96. The average number of documents snowballed per query is 1142, generally smaller than that retrieved from the searched documents list, as the number of seed studies used are smaller.

For the screened-snowballing set, only 34 topics retrieve at least one additional included study. 
The six topics that did not retrieve any additional included studies already contained all included studies prior to snowballing. These topics were 7, 10, 17, 39, 64, and 66. These topics represent an interesting research problem: applying snowballing to these topics would have resulted in wasted time and effort by the researchers of the systematic review. We leave investigations into determining whether or not to apply snowballing for future work.
After removing already screened studies, the screened-snowballing sets contain, on average, 1000 studies. 

\subsubsection{Effectiveness comparison of retrieval methods}

Finally, we aim to investigate the effectiveness of different search methods when used alone and when combined. The results of this analysis are presented in Figure~\ref{figure:precision_recall_boxplot}.
These plots show that using the seed studies alone (i.e., no searching and no snowballing) achieves the lowest recall but the highest precision. Snowballing the seed studies does increase recall but dramatically lowers precision. 
%

Many topics retrieve almost all included studies with the Boolean query. Yet, combining the retrieved studies with other methods further improves recall while lowering precision. Combining both snowballing methods with the retrieved studies obtains the highest recall and lowest precision. However, total recall is still not achieved. Thus, we analysed the recall for different queries in Figure \ref{fig:bar_chart}. 

We found that for some topics, the recall is unusually low. We investigated these topics and found that these systematic reviews used several medical literature databases other than PubMed to retrieve studies. This results in the inclusion of studies that exist in the PubMed database, but are not retrieved by the PubMed query. We also randomly chose some topics reaching total recall and found that even though some of them still use a combination of multiple medical literature databases, they all used PubMed as one of the search sources. Thus, it is possible to create a more effective query for these topics. We leave such a problem for future work.


%% file: Query-Formulation.tex
\vspace{-8pt}
\section{Query Formulation}
We begin our demonstration of the use-cases of our collection with a reproduction of two automatic query formulation methods that use seed studies, see \flabel{1} in Figure~\ref{fig:process}. We use the implementations of~\citet{scells2021comparison}. We compare both automatic query formulation methods when using seed studies and pseudo seed studies.

\label{sec:query-formulation}
\subsection{Methods \& Experimental Settings}
The query formulation experiments are based on two existing methods from the literature~\cite{scells2020automatic,scells2020computational}. These methods are fully automated adaptations of manual or semi-automated procedures that information specialists use in practice. The first is called the \textit{conceptual method} and is what the majority of information specialists use to formulate queries~\cite{suhail2013methods}. The automated conceptual method~\cite{scells2020automatic} takes as input a preliminary string for identifying salient terms (we use the title of the systematic review). The seed studies are then used to optimise the coverage of different combinations of terms expanded from those in the title. The second is called the \textit{objective method} and is a more recent procedure that takes a statistical approach to query formulation~\cite{hausner2012routine}. At a high level, the automated objective method~\cite{scells2020computational} first identifies and ranks salient terms from seed studies using term frequency statistics of the seed studies and a background collection. Next, terms are filtered and added to Boolean clauses by tuning these statistics to a held-out portion of seed studies. Given that the objective method relies on a held-out portion and the conceptual does not, we run both methods for three iterations using different arrangements of seed studies so that both methods use the same set of seed studies each iteration. We run these three iterations twice: once for the real seed studies and once for the pseudo seed studies. 

One other aspect of the automatic versions of the two query formulation methods is the notion of an \textit{instantiation}. In other words, the inclusion or exclusion of different aspects of Boolean queries (e.g., including or excluding MeSH, or including or excluding phrases). To this end, we perform experiments for only the most effective instantiation of each method (Conceptual/Phrase, and Objective/Phrase/Recall/MeSH). We refer the reader to the original study~\cite{scells2020computational} for a comprehensive description of all experimental settings and implementation details that we have used.)

\subsection{Results \& Analysis}
The results of our automatic query formulation reproduction study using our collection are presented in Table~\ref{table:query-formulation}. We report precision, recall, and average number of studies retrieved. 

\begin{table}
	\centering
	\input{results-table-query-formulation.tex}
	\caption{Effectiveness of queries formulated using pseudo seed studies (Pseudo) and seed studies (Seed). Also included are retrieval results of the queries for each topic (Original queries). Oracle experiments and the evaluation measures optimised can be seen in brackets.}
	\label{table:query-formulation}
	\vspace{-24pt}
\end{table}


\subsubsection{Comparison of seed studies and pseudo seed studies}
We first investigate the differences between using seed studies and pseudo seed studies for the objective method.
Using real seed studies to formulate queries is more effective (precision and recall) than using pseudo seed studies for the objective method.
One possible reason seed studies make better queries is that the pseudo seed studies may have been identified through snowballing (and therefore never originally retrieved by terms that exist in the Boolean query).

For the conceptual method, the result indicate that seed studies produce less effective queries than pseudo seed studies. We observe a noticeable decrease in precision and small decrease in recall between the seed studies and pseudo seed studies. However, while all queries using the objective method retrieve at least one study across all three iterations, queries from 17 topics constructed using the conceptual method do not retrieve any studies whatsoever. These results suggest that the automatic conceptual method generally produces less effective queries then the objective method (regardless of the use of seed or pseudo seed studies).

\subsubsection{Comparison of oracle-setting based results}
Given the possible issues with query formulation, and that certain (pseudo) seed studies may impact the effectiveness of resulting queries, we also investigate an oracle approach to identifying the most effective queries across the three iterations. The oracle process can be thought of as simulating the selection of an effective query by choosing from a list of three possible candidate queries.

When using the oracle query, the results show a similar trend as the results obtained from the objective method above: queries constructed using seed studies achieve a higher precision and recall than pseudo seed studies.
The conceptual method shows that even though using seed studies still results in a lower precision, the recall is higher than when pseudo seed studies are used. Again, as 17 topics across all three iterations do not retrieve any studies; this outcome may be biased to those topics that retrieved more than one study.

\subsubsection{Comparison to original queries}
Finally, we compare the results of the two automatic query formulation methods to the retrieval results of the queries for each topic in our collection. The original queries are highly effective, achieving dramatically higher precision and recall results. The average number of studies retrieved is also considerably lower than the automatic methods. We leave the development of more effective queries for systematic review literature search for future work.

\subsection{Overall Findings}

From the automatic query formulation experiments, we found that:

\begin{itemize}[leftmargin=*]
	\item For the objective method, higher effectiveness can be achieved when queries are constructed using seed studies. In fact, pseudo seed studies are detrimental to the query formulation procedure.
	\item For the conceptual method, queries may require manual  modification to prevent biased results. We encountered similar results to~\citet{scells2021comparison} where several topics retrieved no studies. We leave further investixgations for future work.
\end{itemize}

%% file: results-table-query-formulation.tex
\begin{tabular}{p{0.2cm}p{3.2cm}rrr}

	\toprule
	&Method &Precision&Recall&\multirow{2}{1cm}{Avg.\\Retrieved}\\\\
	\hline \addlinespace[2pt]
	
	&Original queries&0.01748&0.73659&1,326 \\ \midrule
	\multirow{6}{*}{\rotatebox{90}{Objective}}&Pseudo &0.00005&0.23457&746,193\\

	&Seed&0.00024&0.31659&806,760\\
	
	\cmidrule{2-5}
	&Pseudo (oracle precision)&0.00015&0.22151&346,023\\
	
	&Seed (oracle precision)&0.00573&0.34188&779,851\\\cmidrule{2-5}
	
	&Pseudo (oracle recall)&0.00014&0.50142&1,550,669\\

	&Seed (oracle recall)&0.00572&0.51923&1,209,792\\
	
	\midrule
	
	\multirow{6}{*}{\rotatebox{90}{Conceptual}}&Pseudo&0.00664&0.27273&433,113\\

	&Seed&0.00093&0.26781&362,968\\ \cmidrule{2-5}

	&Pseudo (oracle precision)&0.00682&0.20940&30,961\\
	
	&Seed (oracle precision)&0.00203&0.29202&360,549\\\cmidrule{2-5}
	
	&Pseudo (oracle recall)&0.00657&0.37334&667,398\\
	&Seed (oracle recall)&0.00183&0.41382&848,223\\ 
	
	\bottomrule
\end{tabular}

%% file: Seed-driven-Document-Ranking.tex
\section{Screening Prioritisation}
\label{sec:sdr}
\begin{table*}[t]
	\centering
	\small
	\input{results-table-sdr-search.tex}
	\caption{Results of baselines and SDR methods on our collection in three experimental settings: single pseudo seed studies, multiple pseudo seed studies, and seed studies. In the header, P refers to `precision' and R refers to `recall'. For AES methods, word2vec PubMed embeddings are denoted by `-P'. AES methods that do not have this demarcation correspond to word2vec embeddings, including PubMed and Wikipedia. Statistical significance (Student's two-tailed paired t-test with Bonferonni correction, $p<0.05$) between SDR method and all other methods is indicated by $\dagger$.}
	\label{table:results.seed_search}
	\vspace{-12pt}
\end{table*}
We continue with another possible use-cases for this collection with a reproduction of a method that uses seed studies for screening prioritisation (i.e., ranking the set of retrieved studies), see \flabel{2} in Figure~\ref{fig:process}. The techniques was originally proposed by~\citet{lee2018seed} and we use the reproduced implementation provided by~\citet{wang2022seed}. We investigate the effectiveness of seed-driven document ranking (SDR) methods, again comparing seed studies with pseudo seed studies. Note that as per previous studies, SDR refers to the specific process of screening prioritisation and a method of screening prioritisation called the SDR method. We make this distinction clear by referring to `SDR' as the task and the `SDR method' as the ranking function.

\subsection{Methods \& Experimental Setup}
The SDR method proposed by~\citet{lee2018seed} is based on the observation that terms in relevant documents are more similar than terms of irrelevant documents. Thus, a ranking model is devised based on this observation which weights each term in the document based on the inter-study similarity. In the SDR method, a study's relevance score is calculated by the sum of pre-computed term weighting multiplied by the likelihood of the term to appear (calculated using the query likelihood model --- QLM). In previous research, two study representation models have been explored:
\begin{description}
	\item[Bag of words] where all terms in a study are used.
	\item[Bag of clinical words] where clinical terms in a study are used.
\end{description} 
Bag of words representations are more effective than the bag of clinical words representation~\cite{wang2022seed}. Therefore, we adopt the bag of words representation (as indicated by \textbf{BOW}) here. Adding to this we investigate a new aspect of SDR: seed studies versus pseudo seed studies.

\subsubsection{Single pseudo seed study}
Firstly, we assume a single included study as an available seed study for each systematic review topic, as per~\citet{lee2018seed}. As there are multiple included studies corresponding to every systematic review topic, the overall effectiveness of retrieval methods on each topic is calculated using the average effectiveness of utilising every included study. Using this leave-one-out cross-validation strategy, the results gathered tend to be more reliable and unbiased across all included studies in a systematic review topic \cite{Leave-One-OutCross-Validation}.\footnote{Topic \#18 only has one included study. We disregard this topic in the evaluation across all experiments.}

\subsubsection{Multiple pseudo seed studies}
Using multiple pseudo seed studies is more effectiveness then using a single seed study for SDR~\cite{wang2022seed}. For our collection, we also perform SDR with multiple pseudo seed studies.
We adopt the experimental settings of~\citet{wang2022seed} to evaluate the effectiveness of SDR when using multiple pseudo seed studies. We adopt the same seed study grouping strategy in which 20\% of included studies are chosen using a sliding-window approach. The groups are then combined by concatenating their titles and abstracts to act as the input to the retrieval methods. The effectiveness on each topic is then calculated using the average effectiveness from all groups.


\subsubsection{Seed studies}
\label{sec:sdr.seed-studies}
Using the seed studies in our collection, we can now realistically investigate the effectiveness of SDR. This experiment combines all the real seed studies by concatenating their titles and abstracts, as used in the multiple pseudo seed study experiments. The combined studies then act as an input for the SDR method and all of the included studies are used for evaluation.

\subsubsection{Retrieval Methods}
Apart from the original SDR method proposed by~\citet{lee2018seed}, we also perform experiments using several baselines: BM25, a query likelihood model (QLM), and a word embedding-based model (AES). As in the previous SDR papers, we used two pre-trained word embeddings for the AES method: one trained on PubMed and Wikipedia and one trained on only PubMed. 
Additionally, we also include fusion methods used in the original paper to interpolate the SDR method with AES using the same parameters in the original article ($\alpha$ = 0.3). 

\subsubsection{Evaluation Measures}
\label{sec:sdr.evaluation-measures}
We evaluate the different arrangements of seed studies and methods with rank-based measures. We use the same evaluation measures as in our reproducibility study~\cite{wang2022seed}. In addition to MAP, which measures the ranking effectiveness of the entire list of studies, we also measure precision, recall, and nDCG at different cut-offs --- \{10,100,1000\}, and Last Relevant\% (LR\%), which reports the percentage of studies that must be screened in order to identify all included studies.

\subsection{Results \& Analysis}

The results of using a single pseudo seed study, multiple pseudo seed studies, and seed studies are shown in Table~\ref{table:results.seed_search}.

\subsubsection{Single and multiple pseudo seed studies}
For single pseudo seed studies, the SDR method improves effectiveness for deep evaluation metrics (i.e., precision@\{100,1000\}, recall@\{100,1000\}, and nDCG@\{100,1000\}). 
SDR-BOW-AES-P is the most effective, apart from shallow evaluation measures (i.e., precision @10, recall@10 and nDCG@10.). For shallow evaluation measures, QLM achieves the highest effectiveness. This may be because the word embeddings based measures help alleviate vocabulary mismatch, improving recall but at the expense of early precision.

For multiple pseudo seed studies, SDR is the most effective on all evaluation measures except precision@1000, recall@1000, and LR\%. Although, it is interesting to observe that the fused method is not able to achieve higher effectiveness than using a single retrieval method alone. This may be due to the poor performance of the AES method, and remains an interesting future challenge for automatically determining when to apply fusion. Using multiple pseudo seed studies is always more effective than single seed studies.

All the aforementioned results are in line with our previous reproduction study~\citet{wang2022seed}, with the exception of the finding here that result fusion did not increase effectiveness.


\subsubsection{Comparison of pseudo-seed results and real-seed results}
\begin{figure}[t]
	\centering
	\includegraphics[width=\columnwidth]{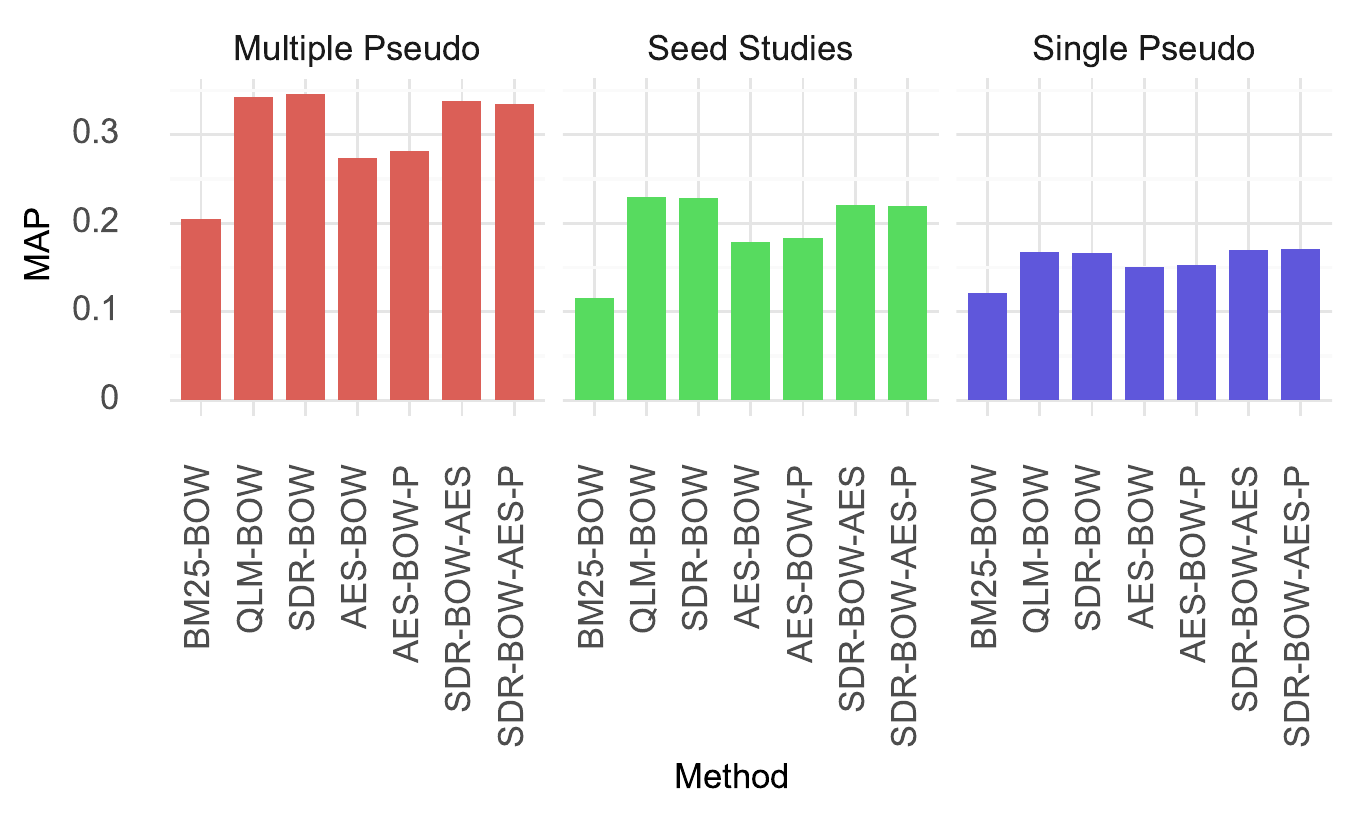}
	\caption{Comparison between the different arrangements of seed and pseudo-seed studies across the SDR methods. We have chosen to show only MAP, but all evaluation measures we chose showed the same trend.}
	\label{fig:sdr-similarity}
	\vspace{-12pt}
\end{figure}

Finally, we found that using real seed studies verses multiple pseudo seed studies dramatically impacts effectiveness.
Firstly, even though the effectiveness of using seed studies still outperforms the use of a single pseudo seed study in some methods, the effectiveness of seed studies is significantly worse than using multiple pseudo seed studies. One possible explanation is that seed studies are not always relevant to the systematic review topic (pseudo seed studies are by definition). Using a non-relevant seed study could degrade the quality of results. This effect is highlighted in Figure~\ref{fig:sdr-similarity}, where the effectiveness of using all seed studies is closer to using a single pseudo study than it is to using multiple pseudo seed studies.

Secondly, using multiple pseudo seed retrieval with the SDR and SDR fused methods consistently outperforms other retrieval methods. However, when using real seed studies, the QLM method is most effective. Our explanation is that while the SDR method still captures the semantic meaning of terms through term weighting, one common use of seed studies in systematic review creation is term extraction, which is better represented by methods like QLM.

In conclusion, pseudo seed studies are not representative of real seed studies and this impacts seed-drive retrieval (SDR). That is why it is important to have test collections with real seed studies such as the one provided in this paper.

\subsection{Overall Findings}

From the SDR experiment, we found that:

\begin{itemize}[leftmargin=*]
	\item Using pseudo seed studies produces unrealistic results compared to using seed studies (i.e., the results are higher than if one used seed studies). This is likely a result of the fact that the pseudo seed studies are an example of explicit relevance feedback, whereas seed studies are more akin to pseudo relevance feedback.
	\item The choice of ranking models in this task (e.g., BM25, QLM, SDR, AES) is dependent on the kind of seed studies used. This may be due to how terms are weighted (i.e., relevant terms are likely to appear in pseudo seed studies but are less likely to appear in seed studies).
\end{itemize}

%% file: results-table-sdr-search.tex
\begin{tabular}{p{0.1cm}l|l|lll|lll|lll|p{0.7cm}}
	
	\toprule
	& Method &MAP&P@10&P@100&P@1000&R@10&R@100&R@1000&nDCG@10&nDCG@100&nDCG@1000&LR\%\\
	\hline \addlinespace[2pt]
	
	\multirow{7}{*}{\rotatebox{90}{Single Pseudo}}&BM25-BOW& 0.1214$^\dagger$&0.1539$^\dagger$&0.0696$^\dagger$&0.0156&0.1226$^\dagger$&0.4049$^\dagger$&0.6855$^\dagger$&0.1846$^\dagger$&0.2760$^\dagger$&0.3764$^\dagger$& 0.6915$^\dagger$\\
	&QLM-BOW& 0.1671& \textbf{0.2190}&0.0829&0.0168& \textbf{0.1646}&0.4572&0.7059& \textbf{0.2636}&0.3377&0.4291&0.5966\\
	&SDR-BOW& 0.1661&0.2165&0.0833&0.0168&0.1632&0.4638&0.7055&0.2599&0.3386&0.4286&0.6117\\
	&AES-BOW& 0.1501&0.1932&0.0807&0.0169&0.1361&0.4349&0.7022&0.2345&0.3141&0.4110&0.6365\\
	&AES-BOW-P& 0.1527&0.1965&0.0831&0.0170&0.1361&0.4455&0.7049&0.2349&0.3184&0.4134&0.6017\\
	&SDR-BOW-AES& 0.1698&0.2185&0.0864&0.0172&0.1573&0.4710&0.7060&0.2615&0.3437&0.4328&0.6007\\
	&SDR-BOW-AES-P&\textbf{0.1709}&0.2165& \textbf{0.0877}& \textbf{0.0172}&0.1531& \textbf{0.4726}& \textbf{0.7114}&0.2596& \textbf{0.3446}& \textbf{0.4337}&\textbf{0.5820}\\
	
	\midrule
	
	\multirow{7}{*}{\rotatebox{90}{Multiple Pseudo}}&BM25-BOW&0.2045$^\dagger$&0.2253$^\dagger$&0.0712$^\dagger$&0.0154&0.1888$^\dagger$&0.4281$^\dagger$&0.6792&0.3145$^\dagger$&0.3817$^\dagger$&0.4758$^\dagger$& 0.7073$^\dagger$\\
	&QLM-BOW& 0.3429&0.4629&0.1222& \textbf{0.0190}&0.2767&0.5531& \textbf{0.7251}&0.5934&0.5520&0.6174&\textbf{0.5255} \\
	&SDR-BOW&\textbf{0.3457}& \textbf{0.4726}& \textbf{0.1231}&0.0189& \textbf{0.2798}& \textbf{0.5630}&0.7226& \textbf{0.6027}& \textbf{0.5574}& \textbf{0.6196}&0.5483\\
	&AES-BOW& 0.2731$^\dagger$&0.3415$^\dagger$&0.1059$^\dagger$&0.0183&0.2228$^\dagger$&0.5041$^\dagger$&0.7147&0.4466$^\dagger$&0.4727$^\dagger$&0.5529$^\dagger$&0.6156$^\dagger$\\
	&AES-BOW-P& 0.2810$^\dagger$&0.3488$^\dagger$&0.1094$^\dagger$&0.0185&0.2260$^\dagger$&0.5202$^\dagger$&0.7179&0.4546$^\dagger$&0.4834$^\dagger$&0.5598$^\dagger$&0.5743\\
	&SDR-BOW-AES& 0.3374&0.4464$^\dagger$&0.1201&0.0189&0.2666&0.5517&0.7211&0.5791$^\dagger$&0.5465$^\dagger$&0.6131&0.5672\\
	&SDR-BOW-AES-P& 0.3344&0.4341$^\dagger$&0.1200&0.0189&0.2634&0.5538&0.7235&0.5662$^\dagger$&0.5439&0.6103&0.5426\\
	
	\midrule
	\multirow{7}{*}{\rotatebox{90}{Seed Studies}}&BM25-BOW& 0.1153$^\dagger$&0.1300$^\dagger$&0.0595$^\dagger$&0.0151&0.1067$^\dagger$&0.3404$^\dagger$&0.6333$^\dagger$&0.1516$^\dagger$&0.2375$^\dagger$&0.3439$^\dagger$& 0.7591$^\dagger$\\
	&QLM-BOW&\textbf{0.2294}& \textbf{0.2850}&0.0932&0.0169&0.2016&0.4614&0.6706& \textbf{0.3370}& \textbf{0.3867}& \textbf{0.4655}&\textbf{0.5749}\\
	&SDR-BOW& 0.2289&0.2800& \textbf{0.0945}&0.0169& \textbf{0.2028}& \textbf{0.4695}&0.6712&0.3253&0.3860&0.4619&0.6012\\
	&AES-BOW& 0.1784$^\dagger$&0.2250&0.0830$^\dagger$&0.0169&0.1644&0.4243&0.6657&0.2570&0.3276$^\dagger$&0.4184$^\dagger$&0.6390\\
	&AES-BOW-P& 0.1835$^\dagger$&0.2275&0.0860&0.0169&0.1614&0.4354&0.6667&0.2656&0.3381$^\dagger$&0.4248$^\dagger$&0.6008\\
	&SDR-BOW-AES& 0.2201&0.2650&0.0927& \textbf{0.0171}&0.1976&0.4615& \textbf{0.6724}&0.3151&0.3790&0.4581&0.5979\\
	&SDR-BOW-AES-P& 0.2198&0.2600&0.0912&0.0169&0.1972&0.4484&0.6680&0.3122&0.3732&0.4549&0.5750\\
	
	\bottomrule
\end{tabular}

%% file: snowballing-baselines.tex
\begin{table*}[t]
	\centering
	\small
	\input{results-table-seed-snowballing-search.tex}
	\caption{Results of SDR methods on our collection when the seed-snowballing set and retrieved studies are combined. Denotations are identical to those in the caption of Table~\ref{table:results.seed_search}.}
	\label{table:results.seed_search_snowball}
	\vspace{-16pt}
\end{table*}
\section{Ranking with Snowballing}
\label{snowballing}
Our last demonstration of the use-cases of our collection is a new technique that we have devised specifically for this paper: ranking with snowballing, see \flabel{3} in Figure~\ref{fig:process}.
Our collection provides two snowballing sets for each topic. Using seed studies and the two snowballing sets, we investigate the impact of snowballing when using SDR.
We investigate two use cases of snowballing within the context of SDR:
\begin{enumerate}[leftmargin=*]
	\item The effectiveness of SDR on a combined set of seed-snowballing studies and retrieved studies.
	\item The effectiveness of SDR on the screened-snowballing set.
\end{enumerate}
These two use cases demonstrate (1) the effectiveness of ranking combined seed-snowballed and retrieved studies sets (i.e., integrating the seed-snowballed set into the retrieved studies and then ranking); and (2) the effectiveness of ranking post-screening (i.e., simulating the ranking of the screened-snowballed set using the screened retrieved studies as input).

\begin{table}[t]
	\centering
	\small
	\input{results-table-screened-snowballing.tex}
	\caption{Results of SDR methods on our collection when using the screened-snowballing set. Denotations are identical to those in the caption of Table~\ref{table:results.seed_search}.}
	\label{table:results.screened_snowball}
	\vspace{-24pt}
\end{table}

\begin{figure}[t]
	\centering
	\includegraphics[width=.66\columnwidth]{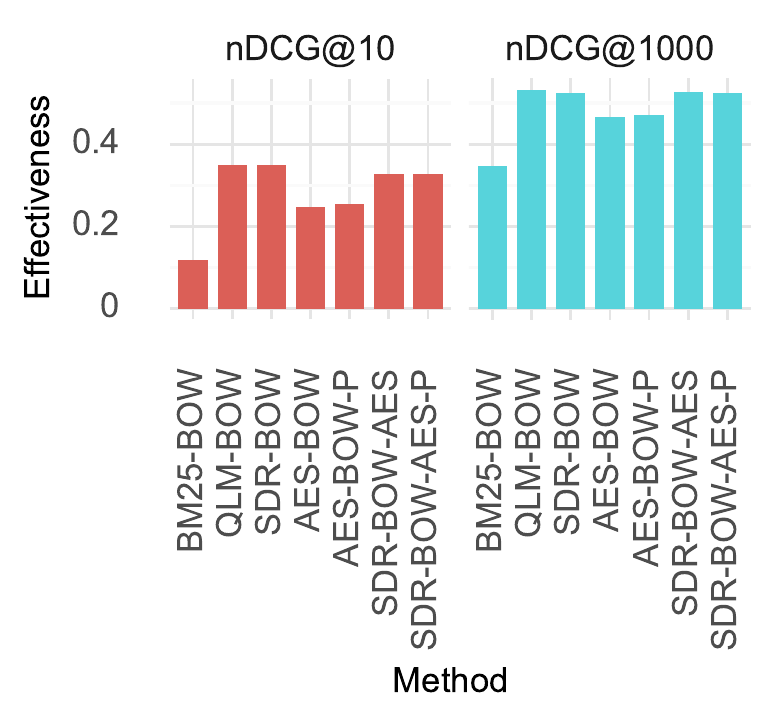}
	\caption{Comparison between nDCG@10 and nDCG@1000 for the results in Table~\ref{table:results.seed_search_snowball}.}
	\label{fig:snowballing-map}
	\vspace{-12pt}
\end{figure}

\begin{figure}[t]
	\centering
	\includegraphics[width=.66\columnwidth]{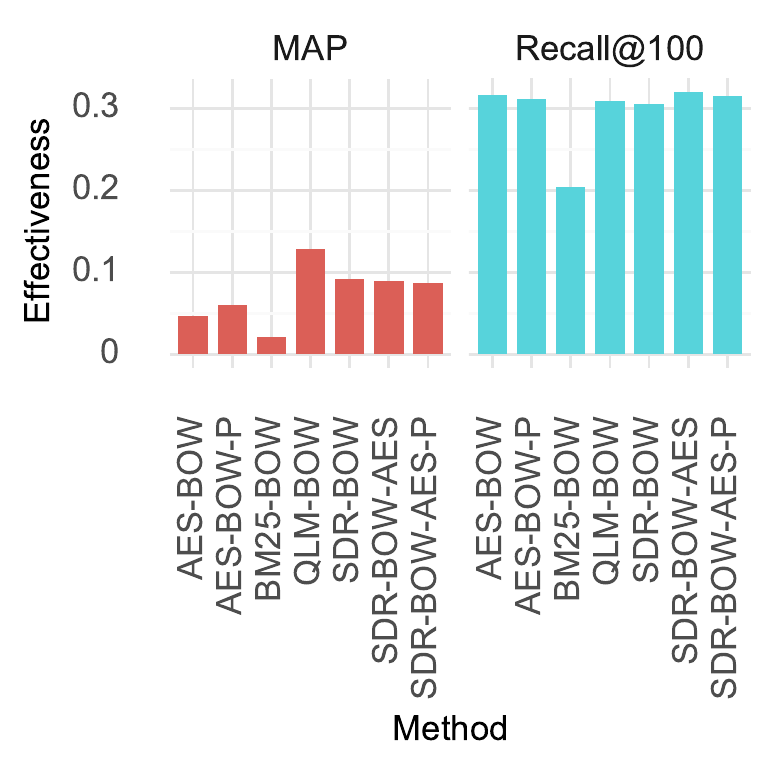}
	\caption{Comparison between MAP and recall@100 for the results in Table~\ref{table:results.screened_snowball}.}
	\label{fig:snowballing-map-recall}
	\vspace{-12pt}
\end{figure}
\subsection{Methods \& Experimental Setup}

\subsubsection{Combined seed-snowballing and retrieved studies ranking}
In Section~\ref{sec:sdr} we investigated the effectiveness of SDR using different arrangements of seed studies. We found that when real seed studies are used, the QLM method outperforms the SDR method in almost all evaluation measures. In this experiment, we investigate if similar results arise when the seed-snowballing set is combined with retrieved studies.
As the first experiment, we used the same experimental setting as in Section~\ref{sec:sdr.seed-studies}. We first combine the retrieved studies and the seed-snowballing set. Next, we concatenate titles and abstracts of seed studies as input to the ranking methods. 
We report the same evaluation measures as described in Section~\ref{sec:sdr.evaluation-measures}.

\subsubsection{Screened snowballing document ranking}

As the second experiment, we simulate the process of screening prioritisation for the screened-snowballed set. Included studies in the retrieved set plus the seed studies are used as input for SDR.
Given the difference in performance of the two sets of studies (i.e., retrieved included studies and seed studies), one could consider different term weighting functions depending on the study. However, we leave such investigation for future work. 
We perform our evaluation on the included studies that did not appear in the retrieved set of studies. This means that the results for this experiment are not comparable to the results of the other screening prioritisation experiments that appear earlier in this paper. We report a subset of the evaluation measures as described in Section~\ref{sec:sdr.evaluation-measures}.

\subsection{Results \& Analysis}

\subsubsection{Combined seed-snowballing and retrieved studies ranking}

Results for this experiment are shown in Table \ref{table:results.seed_search_snowball}.
The most effective methods are QLM and SDR, with SDR more effectiveness on shallow measures and more effectiveness on deeper measures. These results are highlighted further in Figure~\ref{fig:snowballing-map}.
When comparing these combined results with the seed studies result from Table~\ref{table:results.seed_search}, we found that all methods can achieve higher recall\{@100,@1000\} and LR\%, which suggests that adding the seed snowballing set can significantly increase the number of relevant studies retrieved. 

Despite the fact that there are more studies overall when combining the seed-snowballing set with retrieved studies, it is worth it in terms of the overall improvement in effectiveness. In practise, it is beneficial to combine the results of seed-snowballing with retrieved studies for screening prioritisation.

\subsubsection{Screened snowballing document ranking}

Results for this experiment are shown in Table~\ref{table:results.screened_snowball}.
The two best performing methods here were QLM and the AES methods. QLM is the most effective for MAP and nDCG@100. Meanwhile, fusion methods are the most effective for precision@100, recall@100, and LR\%. These results are highlighted in Figure~\ref{fig:snowballing-map-recall}.
These findings demonstrate that QLM is effective for the SDR task when ranking screened-snowballing studies. 

\subsection{Overall Findings}

From the ranking with snowballing experiments, we found that:

\begin{itemize}[leftmargin=*]
	\item Combining the seed-snowballing set and retrieved studies is beneficial for SDR.
	\item QLM and AES are best for screened-snowballing document ranking. However, there is room for future work in determining the optimal combination of studies to use for SDR.
\end{itemize}

%% file: results-table-seed-snowballing-search.tex
\begin{tabular}{p{0.1cm}l|l|lll|lll|lll|p{0.7cm}}
	
	\toprule
	& Method &MAP&P@10&P@100&P@1000&R@10&R@100&R@1000&nDCG@10&nDCG@100&nDCG@1000&LR\%\\
	\hline \addlinespace[2pt]
	
	\multirow{7}{*}{\rotatebox{90}{seed-snowball+search}}&BM25-BOW& 0.0936$^\dagger$&0.0949$^\dagger$&0.0421$^\dagger$&0.0137&0.0943$^\dagger$&0.3372$^\dagger$&0.7767$^\dagger$&0.1175$^\dagger$&0.2074$^\dagger$&0.3469$^\dagger$& 0.7389$^\dagger$\\
	&QLM-BOW&\textbf{0.2672}&0.2744&0.0813&0.0163&0.2453& \textbf{0.5483}&0.8612&0.3490& \textbf{0.4269}& \textbf{0.5326}&\textbf{0.5236}$^\dagger$\\
	&SDR-BOW& 0.2663& \textbf{0.2897}& \textbf{0.0815}&0.0159& \textbf{0.2592}&0.5390&0.8529& \textbf{0.3503}&0.4212&0.5244&0.5672\\
	&AES-BOW& 0.1886$^\dagger$&0.1949$^\dagger$&0.0690& \textbf{0.0165}&0.1815$^\dagger$&0.4619$^\dagger$&0.8595&0.2466$^\dagger$&0.3366$^\dagger$&0.4668$^\dagger$&0.5921\\
	&AES-BOW-P& 0.1950$^\dagger$&0.2026$^\dagger$&0.0700&0.0162&0.1906$^\dagger$&0.4755&0.8530&0.2548$^\dagger$&0.3459$^\dagger$&0.4700$^\dagger$&0.5514\\
	&SDR-BOW-AES& 0.2572&0.2513$^\dagger$&0.0787&0.0164&0.2297$^\dagger$&0.5210& \textbf{0.8675}&0.3271$^\dagger$&0.4136&0.5271&0.5521\\
	&SDR-BOW-AES-P& 0.2546&0.2487$^\dagger$&0.0779&0.0164&0.2343&0.5224&0.8666&0.3266&0.4121&0.5245&0.5298\\
	\bottomrule
\end{tabular}

%% file: results-table-screened-snowballing.tex
\begin{tabular}{p{0.1cm}l|l|lll|p{0.7cm}}
	
	\toprule
	&Method &MAP&Prec&Recall&nDCG&LR\%\\
	&&&@100&@100&@100&\\
	\hline \addlinespace[2pt]
	
	\multirow{7}{*}{\rotatebox{90}{screened-snowball}}&BM25-BOW&0.0208$^\dagger$&0.0100&0.2045&0.0707$^\dagger$&0.4606$^\dagger$\\
	&QLM-BOW&\textbf{0.1279}&0.0212&0.3090&\textbf{0.2104}&0.3317\\
	&SDR-BOW&0.0921&0.0218&0.3055&0.1829&0.3514\\
	&AES-BOW&0.0474&0.0218&0.3167&0.1436&0.3412\\
	&AES-BOW-P&0.0602&0.0206&0.3111&0.1530&0.3329\\

	&SDR-BOW-AES&0.0898&\textbf{0.0226}&\textbf{0.3199}&0.1858&0.3257\\
	&SDR-BOW-AES-P&0.0871&0.0218&0.3149&0.1815&\textbf{0.3214}\\
	\bottomrule
\end{tabular}

%% file: conclusion.tex
\section{Conclusion}
\label{sec:conclusion}

We present a new test collection to properly evaluate systematic review literature search methods which uses seed studies. In addition to our test collection that includes seed studies, we provided a detailed analysis of our collection. Here, we found that as a unique set of studies before query construction, only a small portion of seed studies will be included in the systematic review. We also investigated the impact of seed studies by reproducing two existing methods that use seed studies. Our experiments show that using pseudo seed studies overestimates the effectiveness. 

The test collection also enables an investigation of the difference between snowballing seed studies and screened studies. Here, we found that ranking a combined lists of studies from seed-snowballing documents and searched candidate documents may further boost the effectiveness of ranking models.

The test collection makes two important contributions:
(1) it enables a considerably more realistic evaluation of methods that use seed studies (i.e., one can use all included studies for relevance assessments, instead of requiring a held-out portion as pseudo seed studies); and 
(2) it provides realistic data to develop or train new methods that use seed studies. Seed studies are vital for effective query formulation for information specialists and are commonly used. Despite this, many existing Information Retrieval test collections for systematic review literature search do not contain seed studies. Our test collection will promote the development and realistic evaluation of methods that seed studies can be used to improve systematic review literature search. This includes methods we already explored in this paper like screening prioritisation \cite{lee2018seed, wang2022seed} and query formulation \cite{scells2020automatic}, and some we leave for future works such as active learning \cite{cormack2015autonomy} or MeSH term suggestion \cite{wang_2021_meshterm}, etc. Such methods can have considerable real-world impacts, as systematic reviews are highly time consuming and costly. Cheaper and faster systematic reviews can have dramatic implications for patient outcomes and institutional policy decisions that affect the health decisions of entire countries.

%% file: SR_collection.bbl

\begin{thebibliography}{36}


\ifx \showCODEN    \undefined \def \showCODEN     #1{\unskip}     \fi
\ifx \showDOI      \undefined \def \showDOI       #1{#1}\fi
\ifx \showISBNx    \undefined \def \showISBNx     #1{\unskip}     \fi
\ifx \showISBNxiii \undefined \def \showISBNxiii  #1{\unskip}     \fi
\ifx \showISSN     \undefined \def \showISSN      #1{\unskip}     \fi
\ifx \showLCCN     \undefined \def \showLCCN      #1{\unskip}     \fi
\ifx \shownote     \undefined \def \shownote      #1{#1}          \fi
\ifx \showarticletitle \undefined \def \showarticletitle #1{#1}   \fi
\ifx \showURL      \undefined \def \showURL       {\relax}        \fi
\providecommand\bibfield[2]{#2}
\providecommand\bibinfo[2]{#2}
\providecommand\natexlab[1]{#1}
\providecommand\showeprint[2][]{arXiv:#2}

\bibitem[\protect\citeauthoryear{Abualsaud, Ghelani, Zhang, Smucker, Cormack,
  and Grossman}{Abualsaud et~al\mbox{.}}{2018}]%
        {abualsaud2018system}
\bibfield{author}{\bibinfo{person}{Mustafa Abualsaud}, \bibinfo{person}{Nimesh
  Ghelani}, \bibinfo{person}{Haotian Zhang}, \bibinfo{person}{Mark~D Smucker},
  \bibinfo{person}{Gordon~V Cormack}, {and} \bibinfo{person}{Maura~R
  Grossman}.} \bibinfo{year}{2018}\natexlab{}.
\newblock \showarticletitle{A system for efficient high-recall retrieval}. In
  \bibinfo{booktitle}{\emph{The 41st international ACM SIGIR conference on
  research \& development in information retrieval}}.
  \bibinfo{pages}{1317--1320}.
\newblock


\bibitem[\protect\citeauthoryear{Alharbi and Stevenson}{Alharbi and
  Stevenson}{2019}]%
        {alharbi2019dataset}
\bibfield{author}{\bibinfo{person}{Amal Alharbi} {and} \bibinfo{person}{Mark
  Stevenson}.} \bibinfo{year}{2019}\natexlab{}.
\newblock \showarticletitle{A dataset of systematic review updates}. In
  \bibinfo{booktitle}{\emph{Proceedings of the 42nd International ACM SIGIR
  Conference on Research and Development in Information Retrieval}}.
  \bibinfo{pages}{1257--1260}.
\newblock


\bibitem[\protect\citeauthoryear{Bullers, Howard, Hanson, Kearns, Orriola,
  Polo, and Sakmar}{Bullers et~al\mbox{.}}{2018}]%
        {Bullers:2018vc}
\bibfield{author}{\bibinfo{person}{Krystal Bullers}, \bibinfo{person}{Allison~M
  Howard}, \bibinfo{person}{Ardis Hanson}, \bibinfo{person}{William~D Kearns},
  \bibinfo{person}{John~J Orriola}, \bibinfo{person}{Randall~L Polo}, {and}
  \bibinfo{person}{Kristen~A Sakmar}.} \bibinfo{year}{2018}\natexlab{}.
\newblock \showarticletitle{It Takes Longer than You Think: Librarian Time
  Spent on Systematic Review Tasks}.
\newblock \bibinfo{journal}{\emph{Journal of the Medical Library Association}}
  \bibinfo{volume}{106}, \bibinfo{number}{2} (\bibinfo{year}{2018}),
  \bibinfo{pages}{198}.
\newblock


\bibitem[\protect\citeauthoryear{Clark}{Clark}{2013}]%
        {suhail2013methods}
\bibfield{author}{\bibinfo{person}{Justin Clark}.}
  \bibinfo{year}{2013}\natexlab{}.
\newblock \showarticletitle{Systematic Reviewing}.
\newblock In \bibinfo{booktitle}{\emph{Methods of Clinical Epidemiology}},
  \bibfield{editor}{\bibinfo{person}{Gail M.~Williams Suhail A. R.~Doi}} (Ed.).
\newblock


\bibitem[\protect\citeauthoryear{Clark, Glasziou, {Del Mar}, Bannach-Brown,
  Stehlik, and Scott}{Clark et~al\mbox{.}}{2020}]%
        {2-week-sys-review}
\bibfield{author}{\bibinfo{person}{Justin Clark}, \bibinfo{person}{Paul
  Glasziou}, \bibinfo{person}{Chris {Del Mar}}, \bibinfo{person}{Alexandra
  Bannach-Brown}, \bibinfo{person}{Paulina Stehlik}, {and}
  \bibinfo{person}{{Anna Mae} Scott}.} \bibinfo{year}{2020}\natexlab{}.
\newblock \showarticletitle{A full systematic review was completed in 2 weeks
  using automation tools: a case study}.
\newblock \bibinfo{journal}{\emph{Journal of Chronic Diseases}}
  \bibinfo{volume}{121} (\bibinfo{date}{May} \bibinfo{year}{2020}),
  \bibinfo{pages}{81--90}.
\newblock
\showISSN{0895-4356}
\urldef\tempurl%
\url{https://doi.org/10.1016/j.jclinepi.2020.01.008}
\showDOI{\tempurl}
\newblock
\shownote{Copyright {\textcopyright} 2020 Elsevier Inc. All rights reserved.}


\bibitem[\protect\citeauthoryear{Colace, Santo, Greco, and Napoletano}{Colace
  et~al\mbox{.}}{2011}]%
        {colace2011improving}
\bibfield{author}{\bibinfo{person}{Francesco Colace},
  \bibinfo{person}{Massimo~De Santo}, \bibinfo{person}{Luca Greco}, {and}
  \bibinfo{person}{Paolo Napoletano}.} \bibinfo{year}{2011}\natexlab{}.
\newblock \showarticletitle{Improving text retrieval accuracy by using a
  minimal relevance feedback}. In \bibinfo{booktitle}{\emph{International Joint
  Conference on Knowledge Discovery, Knowledge Engineering, and Knowledge
  Management}}. Springer, \bibinfo{pages}{126--140}.
\newblock


\bibitem[\protect\citeauthoryear{Cormack and Grossman}{Cormack and
  Grossman}{2014}]%
        {cormack2014evaluation}
\bibfield{author}{\bibinfo{person}{Gordon~V Cormack} {and}
  \bibinfo{person}{Maura~R Grossman}.} \bibinfo{year}{2014}\natexlab{}.
\newblock \showarticletitle{Evaluation of machine-learning protocols for
  technology-assisted review in electronic discovery}. In
  \bibinfo{booktitle}{\emph{Proceedings of the 37th international ACM SIGIR
  conference on Research \& development in information retrieval}}.
  \bibinfo{pages}{153--162}.
\newblock


\bibitem[\protect\citeauthoryear{Cormack and Grossman}{Cormack and
  Grossman}{2015}]%
        {cormack2015autonomy}
\bibfield{author}{\bibinfo{person}{Gordon~V Cormack} {and}
  \bibinfo{person}{Maura~R Grossman}.} \bibinfo{year}{2015}\natexlab{}.
\newblock \showarticletitle{Autonomy and reliability of continuous active
  learning for technology-assisted review}.
\newblock \bibinfo{journal}{\emph{arXiv preprint arXiv:1504.06868}}
  (\bibinfo{year}{2015}).
\newblock


\bibitem[\protect\citeauthoryear{Cormack and Grossman}{Cormack and
  Grossman}{2016}]%
        {cormack2016scalability}
\bibfield{author}{\bibinfo{person}{Gordon~V Cormack} {and}
  \bibinfo{person}{Maura~R Grossman}.} \bibinfo{year}{2016}\natexlab{}.
\newblock \showarticletitle{Scalability of continuous active learning for
  reliable high-recall text classification}. In
  \bibinfo{booktitle}{\emph{Proceedings of the 25th ACM international on
  conference on information and knowledge management}}.
  \bibinfo{pages}{1039--1048}.
\newblock


\bibitem[\protect\citeauthoryear{Greenhalgh and Peacock}{Greenhalgh and
  Peacock}{2005}]%
        {greenhalgh2005effectiveness}
\bibfield{author}{\bibinfo{person}{Trisha Greenhalgh} {and}
  \bibinfo{person}{Richard Peacock}.} \bibinfo{year}{2005}\natexlab{}.
\newblock \showarticletitle{Effectiveness and efficiency of search methods in
  systematic reviews of complex evidence: audit of primary sources}.
\newblock \bibinfo{journal}{\emph{Bmj}} \bibinfo{volume}{331},
  \bibinfo{number}{7524} (\bibinfo{year}{2005}), \bibinfo{pages}{1064--1065}.
\newblock


\bibitem[\protect\citeauthoryear{Haddaway, Grainger, and Gray}{Haddaway
  et~al\mbox{.}}{2021}]%
        {haddaway_n_r_2021_4543513}
\bibfield{author}{\bibinfo{person}{N~R Haddaway}, \bibinfo{person}{M.~J.
  Grainger}, {and} \bibinfo{person}{C.~T. Gray}.}
  \bibinfo{year}{2021}\natexlab{}.
\newblock \bibinfo{booktitle}{\emph{{citationchaser: An R package and Shiny app
  for forward and backward citations chasing in academic searching}}}.
\newblock
\urldef\tempurl%
\url{https://doi.org/10.5281/zenodo.4543513}
\showDOI{\tempurl}


\bibitem[\protect\citeauthoryear{Hausner, Waffenschmidt, Kaiser, and
  Simon}{Hausner et~al\mbox{.}}{2012}]%
        {hausner2012routine}
\bibfield{author}{\bibinfo{person}{Elke Hausner}, \bibinfo{person}{Siw
  Waffenschmidt}, \bibinfo{person}{Thomas Kaiser}, {and}
  \bibinfo{person}{Michael Simon}.} \bibinfo{year}{2012}\natexlab{}.
\newblock \showarticletitle{Routine Development of Objectively Derived Search
  Strategies}.
\newblock \bibinfo{journal}{\emph{Systematic reviews}} \bibinfo{volume}{1},
  \bibinfo{number}{1} (\bibinfo{year}{2012}), \bibinfo{pages}{19}.
\newblock


\bibitem[\protect\citeauthoryear{Jayarathna, Patra, and Shipman}{Jayarathna
  et~al\mbox{.}}{2015}]%
        {jayarathna2015unified}
\bibfield{author}{\bibinfo{person}{Sampath Jayarathna}, \bibinfo{person}{Atish
  Patra}, {and} \bibinfo{person}{Frank Shipman}.}
  \bibinfo{year}{2015}\natexlab{}.
\newblock \showarticletitle{Unified relevance feedback for multi-application
  user interest modeling}. In \bibinfo{booktitle}{\emph{Proceedings of the 15th
  ACM/IEEE-CS Joint Conference on Digital Libraries}}.
  \bibinfo{pages}{129--138}.
\newblock


\bibitem[\protect\citeauthoryear{Kanoulas, Li, Azzopardi, and Spijker}{Kanoulas
  et~al\mbox{.}}{2017}]%
        {kanoulas2017clef}
\bibfield{author}{\bibinfo{person}{Evangelos Kanoulas}, \bibinfo{person}{Dan
  Li}, \bibinfo{person}{Leif Azzopardi}, {and} \bibinfo{person}{Rene Spijker}.}
  \bibinfo{year}{2017}\natexlab{}.
\newblock \showarticletitle{CLEF 2017 technologically assisted reviews in
  empirical medicine overview}. In \bibinfo{booktitle}{\emph{CEUR workshop
  proceedings}}, Vol.~\bibinfo{volume}{1866}. \bibinfo{pages}{1--29}.
\newblock


\bibitem[\protect\citeauthoryear{Kanoulas, Li, Azzopardi, and Spijker}{Kanoulas
  et~al\mbox{.}}{2018}]%
        {Kanoulas2018CLEF2T}
\bibfield{author}{\bibinfo{person}{E. Kanoulas}, \bibinfo{person}{Dan Li},
  \bibinfo{person}{Leif Azzopardi}, {and} \bibinfo{person}{Ren{\'e} Spijker}.}
  \bibinfo{year}{2018}\natexlab{}.
\newblock \showarticletitle{CLEF 2018 Technologically Assisted Reviews in
  Empirical Medicine Overview}. In \bibinfo{booktitle}{\emph{CLEF}}.
\newblock


\bibitem[\protect\citeauthoryear{Kanoulas, Li, Azzopardi, and Spijker}{Kanoulas
  et~al\mbox{.}}{2019}]%
        {kanoulas2019clef}
\bibfield{author}{\bibinfo{person}{Evangelos Kanoulas}, \bibinfo{person}{Dan
  Li}, \bibinfo{person}{Leif Azzopardi}, {and} \bibinfo{person}{Rene Spijker}.}
  \bibinfo{year}{2019}\natexlab{}.
\newblock \showarticletitle{CLEF 2019 technology assisted reviews in empirical
  medicine overview}. In \bibinfo{booktitle}{\emph{CEUR workshop proceedings}},
  Vol.~\bibinfo{volume}{2380}.
\newblock


\bibitem[\protect\citeauthoryear{Kaptein, Kamps, and Hiemstra}{Kaptein
  et~al\mbox{.}}{2008}]%
        {kaptein2008impact}
\bibfield{author}{\bibinfo{person}{Rianne Kaptein}, \bibinfo{person}{Jaap
  Kamps}, {and} \bibinfo{person}{Djoerd Hiemstra}.}
  \bibinfo{year}{2008}\natexlab{}.
\newblock \bibinfo{booktitle}{\emph{The impact of positive, negative and
  topical relevance feedback}}.
\newblock \bibinfo{type}{{T}echnical {R}eport}. \bibinfo{institution}{AMSTERDAM
  UNIV (NETHERLANDS)}.
\newblock


\bibitem[\protect\citeauthoryear{Lagopoulos, Anagnostou, Minas, and
  Tsoumakas}{Lagopoulos et~al\mbox{.}}{2018}]%
        {lagopoulos2018learning}
\bibfield{author}{\bibinfo{person}{Athanasios Lagopoulos},
  \bibinfo{person}{Antonios Anagnostou}, \bibinfo{person}{Adamantios Minas},
  {and} \bibinfo{person}{Grigorios Tsoumakas}.}
  \bibinfo{year}{2018}\natexlab{}.
\newblock \showarticletitle{Learning-to-rank and relevance feedback for
  literature appraisal in empirical medicine}. In
  \bibinfo{booktitle}{\emph{International conference of the cross-language
  evaluation forum for European languages}}. Springer, \bibinfo{pages}{52--63}.
\newblock


\bibitem[\protect\citeauthoryear{Lavrenko and Croft}{Lavrenko and
  Croft}{2017}]%
        {lavrenko2017relevance}
\bibfield{author}{\bibinfo{person}{Victor Lavrenko} {and}
  \bibinfo{person}{W~Bruce Croft}.} \bibinfo{year}{2017}\natexlab{}.
\newblock \showarticletitle{Relevance-based language models}. In
  \bibinfo{booktitle}{\emph{ACM SIGIR Forum}}, Vol.~\bibinfo{volume}{51}. ACM
  New York, NY, USA, \bibinfo{pages}{260--267}.
\newblock


\bibitem[\protect\citeauthoryear{Lee and Sun}{Lee and Sun}{2018}]%
        {lee2018seed}
\bibfield{author}{\bibinfo{person}{Grace~E Lee} {and} \bibinfo{person}{Aixin
  Sun}.} \bibinfo{year}{2018}\natexlab{}.
\newblock \showarticletitle{Seed-driven document ranking for systematic reviews
  in evidence-based medicine}. In \bibinfo{booktitle}{\emph{The 41st
  international ACM SIGIR conference on research \& development in information
  retrieval}}. \bibinfo{pages}{455--464}.
\newblock


\bibitem[\protect\citeauthoryear{Maglott, Ostell, Pruitt, and Tatusova}{Maglott
  et~al\mbox{.}}{2005}]%
        {maglott2005entrez}
\bibfield{author}{\bibinfo{person}{Donna Maglott}, \bibinfo{person}{Jim
  Ostell}, \bibinfo{person}{Kim~D Pruitt}, {and} \bibinfo{person}{Tatiana
  Tatusova}.} \bibinfo{year}{2005}\natexlab{}.
\newblock \showarticletitle{Entrez Gene: gene-centered information at NCBI}.
\newblock \bibinfo{journal}{\emph{Nucleic acids research}}
  \bibinfo{volume}{33}, \bibinfo{number}{suppl\_1} (\bibinfo{year}{2005}),
  \bibinfo{pages}{D54--D58}.
\newblock


\bibitem[\protect\citeauthoryear{McGowan and Sampson}{McGowan and
  Sampson}{2005}]%
        {McGowan:2005up}
\bibfield{author}{\bibinfo{person}{Jessie McGowan} {and}
  \bibinfo{person}{Margaret Sampson}.} \bibinfo{year}{2005}\natexlab{}.
\newblock \showarticletitle{Systematic Reviews Need Systematic Searchers
  ({{IRP}})}.
\newblock \bibinfo{journal}{\emph{Journal of the Medical Library Association}}
  \bibinfo{volume}{93}, \bibinfo{number}{1} (\bibinfo{year}{2005}),
  \bibinfo{pages}{74}.
\newblock


\bibitem[\protect\citeauthoryear{Miew Keen~Choong}{Miew Keen~Choong}{2014}]%
        {snowballing-definition-choong}
\bibfield{author}{\bibinfo{person}{Adam G Dunn Guy~Tsafnat Miew Keen~Choong,
  Filippo~Galgani}.} \bibinfo{year}{2014}\natexlab{}.
\newblock \showarticletitle{Automatic Evidence Retrieval for Systematic
  Reviews}.
\newblock \bibinfo{journal}{\emph{J Med Internet Res}} (\bibinfo{date}{10}
  \bibinfo{year}{2014}).
\newblock


\bibitem[\protect\citeauthoryear{Norman, Leeflang, and N{\'e}v{\'e}ol}{Norman
  et~al\mbox{.}}{2018}]%
        {norman2018data}
\bibfield{author}{\bibinfo{person}{Christopher Norman},
  \bibinfo{person}{Mariska Leeflang}, {and} \bibinfo{person}{Aur{\'e}lie
  N{\'e}v{\'e}ol}.} \bibinfo{year}{2018}\natexlab{}.
\newblock \showarticletitle{Data extraction and synthesis in systematic reviews
  of diagnostic test accuracy: a corpus for automating and evaluating the
  process}. In \bibinfo{booktitle}{\emph{AMIA Annual Symposium Proceedings}},
  Vol.~\bibinfo{volume}{2018}. American Medical Informatics Association,
  \bibinfo{pages}{817}.
\newblock


\bibitem[\protect\citeauthoryear{Pereira, Etemad, and Paulovich}{Pereira
  et~al\mbox{.}}{2020}]%
        {pereira2020iterative}
\bibfield{author}{\bibinfo{person}{Mateus Pereira}, \bibinfo{person}{Elham
  Etemad}, {and} \bibinfo{person}{Fernando Paulovich}.}
  \bibinfo{year}{2020}\natexlab{}.
\newblock \showarticletitle{Iterative learning to rank from explicit relevance
  feedback}. In \bibinfo{booktitle}{\emph{Proceedings of the 35th Annual ACM
  Symposium on Applied Computing}}. \bibinfo{pages}{698--705}.
\newblock


\bibitem[\protect\citeauthoryear{Prasad, Kaur, and Kan}{Prasad
  et~al\mbox{.}}{2018}]%
        {animesh2018neuralparscit}
\bibfield{author}{\bibinfo{person}{Animesh Prasad}, \bibinfo{person}{Manpreet
  Kaur}, {and} \bibinfo{person}{Min-Yen Kan}.} \bibinfo{year}{2018}\natexlab{}.
\newblock \showarticletitle{Neural {{ParsCit}}: {{A}} Deep Learning Based
  Reference String Parser}.
\newblock \bibinfo{journal}{\emph{Journal on Digital Libraries}}
  \bibinfo{volume}{19} (\bibinfo{year}{2018}), \bibinfo{pages}{323--337}.
\newblock


\bibitem[\protect\citeauthoryear{Rocchio and Salton}{Rocchio and
  Salton}{1965}]%
        {rocchio1965information}
\bibfield{author}{\bibinfo{person}{JJ Rocchio} {and} \bibinfo{person}{Gerard
  Salton}.} \bibinfo{year}{1965}\natexlab{}.
\newblock \showarticletitle{Information search optimization and interactive
  retrieval techniques}. In \bibinfo{booktitle}{\emph{Proceedings of the
  November 30--December 1, 1965, fall joint computer conference, part I}}.
  \bibinfo{pages}{293--305}.
\newblock


\bibitem[\protect\citeauthoryear{Sammut and Webb}{Sammut and Webb}{2010}]%
        {Leave-One-OutCross-Validation}
\bibfield{editor}{\bibinfo{person}{Claude Sammut} {and}
  \bibinfo{person}{Geoffrey~I. Webb}} (Eds.). \bibinfo{year}{2010}\natexlab{}.
\newblock \bibinfo{booktitle}{\emph{Leave-One-Out Cross-Validation}}.
\newblock \bibinfo{publisher}{Springer US}, \bibinfo{address}{Boston, MA},
  \bibinfo{pages}{600--601}.
\newblock
\showISBNx{978-0-387-30164-8}
\urldef\tempurl%
\url{https://doi.org/10.1007/978-0-387-30164-8_469}
\showDOI{\tempurl}


\bibitem[\protect\citeauthoryear{Scells, Zuccon, and Koopman}{Scells
  et~al\mbox{.}}{2021}]%
        {scells2021comparison}
\bibfield{author}{\bibinfo{person}{Harrisen Scells}, \bibinfo{person}{Guido
  Zuccon}, {and} \bibinfo{person}{Bevan Koopman}.}
  \bibinfo{year}{2021}\natexlab{}.
\newblock \showarticletitle{A comparison of automatic Boolean query formulation
  for systematic reviews}.
\newblock \bibinfo{journal}{\emph{Information Retrieval Journal}}
  \bibinfo{volume}{24}, \bibinfo{number}{1} (\bibinfo{year}{2021}),
  \bibinfo{pages}{3--28}.
\newblock


\bibitem[\protect\citeauthoryear{Scells, Zuccon, Koopman, and Clark}{Scells
  et~al\mbox{.}}{2020a}]%
        {scells2020automatic}
\bibfield{author}{\bibinfo{person}{Harrisen Scells}, \bibinfo{person}{Guido
  Zuccon}, \bibinfo{person}{Bevan Koopman}, {and} \bibinfo{person}{Justin
  Clark}.} \bibinfo{year}{2020}\natexlab{a}.
\newblock \showarticletitle{Automatic Boolean query formulation for systematic
  review literature search}. In \bibinfo{booktitle}{\emph{Proceedings of The
  Web Conference 2020}}. \bibinfo{pages}{1071--1081}.
\newblock


\bibitem[\protect\citeauthoryear{Scells, Zuccon, Koopman, and Clark}{Scells
  et~al\mbox{.}}{2020b}]%
        {scells2020computational}
\bibfield{author}{\bibinfo{person}{Harrisen Scells}, \bibinfo{person}{Guido
  Zuccon}, \bibinfo{person}{Bevan Koopman}, {and} \bibinfo{person}{Justin
  Clark}.} \bibinfo{year}{2020}\natexlab{b}.
\newblock \showarticletitle{A computational approach for objectively derived
  systematic review search strategies}. In \bibinfo{booktitle}{\emph{European
  conference on information retrieval}}. Springer, \bibinfo{pages}{385--398}.
\newblock


\bibitem[\protect\citeauthoryear{Scells, Zuccon, Koopman, Deacon, Azzopardi,
  and Geva}{Scells et~al\mbox{.}}{2017}]%
        {scells2017test}
\bibfield{author}{\bibinfo{person}{Harrisen Scells}, \bibinfo{person}{Guido
  Zuccon}, \bibinfo{person}{Bevan Koopman}, \bibinfo{person}{Anthony Deacon},
  \bibinfo{person}{Leif Azzopardi}, {and} \bibinfo{person}{Shlomo Geva}.}
  \bibinfo{year}{2017}\natexlab{}.
\newblock \showarticletitle{A test collection for evaluating retrieval of
  studies for inclusion in systematic reviews}. In
  \bibinfo{booktitle}{\emph{Proceedings of the 40th International ACM SIGIR
  Conference on Research and Development in Information Retrieval}}.
  \bibinfo{pages}{1237--1240}.
\newblock


\bibitem[\protect\citeauthoryear{Wang, Li, Scells, Locke, and Zuccon}{Wang
  et~al\mbox{.}}{2021}]%
        {wang_2021_meshterm}
\bibfield{author}{\bibinfo{person}{Shuai Wang}, \bibinfo{person}{Hang Li},
  \bibinfo{person}{Harrisen Scells}, \bibinfo{person}{Daniel Locke}, {and}
  \bibinfo{person}{Guido Zuccon}.} \bibinfo{year}{2021}\natexlab{}.
\newblock \showarticletitle{MeSH Term Suggestion for Systematic Review
  Literature Search}. In \bibinfo{booktitle}{\emph{Proceedings of the 25th
  Australasian Document Computing Symposium}} (Virtual Event, Australia)
  \emph{(\bibinfo{series}{ADCS '21})}. \bibinfo{publisher}{Association for
  Computing Machinery}, \bibinfo{address}{New York, NY, USA}, Article
  \bibinfo{articleno}{8}, \bibinfo{numpages}{8}~pages.
\newblock
\showISBNx{9781450395991}
\urldef\tempurl%
\url{https://doi.org/10.1145/3503516.3503530}
\showDOI{\tempurl}


\bibitem[\protect\citeauthoryear{Wang, Scells, Mourad, and Zuccon}{Wang
  et~al\mbox{.}}{2022}]%
        {wang2022seed}
\bibfield{author}{\bibinfo{person}{Shuai Wang}, \bibinfo{person}{Harrisen
  Scells}, \bibinfo{person}{Ahmed Mourad}, {and} \bibinfo{person}{Guido
  Zuccon}.} \bibinfo{year}{2022}\natexlab{}.
\newblock \showarticletitle{Seed-Driven Document Ranking for Systematic
  Reviews: A Reproducibility Study}. In \bibinfo{booktitle}{\emph{European
  Conference on Information Retrieval}}. Springer, \bibinfo{pages}{686--700}.
\newblock


\bibitem[\protect\citeauthoryear{Zhang, Abualsaud, Ghelani, Smucker, Cormack,
  and Grossman}{Zhang et~al\mbox{.}}{2018}]%
        {zhang2018effective}
\bibfield{author}{\bibinfo{person}{Haotian Zhang}, \bibinfo{person}{Mustafa
  Abualsaud}, \bibinfo{person}{Nimesh Ghelani}, \bibinfo{person}{Mark~D
  Smucker}, \bibinfo{person}{Gordon~V Cormack}, {and} \bibinfo{person}{Maura~R
  Grossman}.} \bibinfo{year}{2018}\natexlab{}.
\newblock \showarticletitle{Effective user interaction for high-recall
  retrieval: Less is more}. In \bibinfo{booktitle}{\emph{Proceedings of the
  27th ACM international conference on information and knowledge management}}.
  \bibinfo{pages}{187--196}.
\newblock


\bibitem[\protect\citeauthoryear{Zou, Li, and Kanoulas}{Zou
  et~al\mbox{.}}{2018}]%
        {zou2018technology}
\bibfield{author}{\bibinfo{person}{Jie Zou}, \bibinfo{person}{Dan Li}, {and}
  \bibinfo{person}{Evangelos Kanoulas}.} \bibinfo{year}{2018}\natexlab{}.
\newblock \showarticletitle{Technology assisted reviews: Finding the last few
  relevant documents by asking yes/no questions to reviewers}. In
  \bibinfo{booktitle}{\emph{The 41st International ACM SIGIR Conference on
  Research \& Development in Information Retrieval}}.
  \bibinfo{pages}{949--952}.
\newblock


\end{thebibliography}
